\documentclass[a4paper,aps,pra,showpacs,preprintnumbers,twocolumn]{revtex4}
\usepackage[cp1251]{inputenc}
\usepackage[english]{babel}
\usepackage{amssymb,amsfonts,amsmath,mathtext,enumerate,float,dsfont}
\usepackage{graphics,graphicx,epsfig,epstopdf}
\usepackage{caption}
\usepackage{indentfirst}
\usepackage[usenames]{color}
\usepackage{amsthm}

\renewcommand{\Re}{\mathop{\mathrm{Re}}\nolimits}

\begin{document}
\title{Finding the optimal cluster state configuration. Cluster states classification by type of computations.}
\author{Korolev S. B.}
\email{Sergey.Koroleev@gmail.com}
\author{Golubeva T. Yu.}
\author{Golubev Yu. M.}
\affiliation{St. Petersburg State University, Universitetskaya  nab. 7/9, 199034 St. Petersburg, Russia}
\date{\today}
\begin{abstract}
In this paper, we study the transformations that are obtained in one-way quantum computation on continuous-variable cluster states of various configurations. Of all possible cluster configurations, we choose those that are suitable for universal Gaussian operations.
\end{abstract}
\maketitle

\section{Introduction}
One-way quantum computations (OWQC) with continuous-variables are a convenient alternative to computations performed with qubits \cite{Raus2}. All OWQC are realized through successive local measurements of a multipartite entangled state called a cluster state. However, despite the general principle of construction of computation, discrete and continuous systems are significantly different from each other.  This distinction is based on the various properties of the quantum states of the physical systems that compose cluster nodes.

Physical systems in the quadrature-squeezed state (squeezed quantum oscillators) are used to generate continuous-variable clusters.  Each such oscillator is defined by two quadratures $ \hat x_s $ and $ \hat y_s $, which obey the canonical commutation relation $ \left [\hat x_s, \hat y_s \right] = i / 2 $.  The squeezing of oscillators means that the variance in one of the quadratures is smaller than the variance of the vacuum state. It is generally accepted to consider oscillators squeezed in $\hat y_s$-quadrature,  $\langle \delta \hat y_s ^2\rangle \textless {1}/{4}$.

The Gaussian nature of cluster nodes defines a set of operations that transform an arbitrary input quantum state into an output one in a unitary manner. If the input states are encoded on physical systems in continuous variables then to perform universal transformations a quantum computer should be able to implement three types of operations: arbitrary single-mode Gaussian operations, one any two-mode Gaussian operation (usually, \emph{CZ} is considered as a two-mode operation ) and one single-mode non-Gaussian operation \cite{Lloyd}.  The first two operations are generators of the Gaussian transformations group. In other words, any Gaussian transformation can be obtained using these two operations.  By definition,  the Gaussian transformation is the canonical transformation mapping linearly the input quadratures to output.

Thus, if we know the procedure we need to perform, we will be able to evaluate cluster state configuration to implement it. It should be noted here that the question of the requirements for the cluster state configuration is still open.  To date, one shows the feasibility of certain operations on clusters of specific configurations.  For example, \cite{Ukai} demonstrates the implementation of single-mode operation on a linear four-mode cluster, and \cite{OPO} shows that the pair of two-mode clusters can be used to perform such a procedure. The authors of \cite{Su} demonstrate a two-mode operation when input states mix with the two external nodes of a linear cluster. Authors of paper \cite{Korolev_1} discuss the realization of the same operation on an ensemble of two-node cluster states. However, the question of whether any of the above cluster configurations are preferred remains open.

In this work, we analyze all possible cluster configurations for the feasibility of universal Gaussian transformations on them. It is well-known \cite{Ukai, OPO, Su} that the result of such transformations can be written in vector form:
 \begin{align} \label{1}
    \begin{pmatrix}
        \hat {\vec{X}}_{out}\\
        \hat {\vec{Y}}_{out}
    \end{pmatrix} = \tilde {U}    \begin{pmatrix}
        \hat {\vec{x}}_{in}\\
        \hat {\vec{y}}_{in}
    \end{pmatrix}+E \hat{\vec{y}}_s,
 \end{align}
where $\tilde U$  is a main transformation matrix, $E$ is an error matrix, $\hat {\vec{y}}_s$ -- is a vector consisting of squeezed $\hat {y}$ - quadratures of all cluster nodes,  $\hat{\vec{x}}_{in}$ and $ \hat{\vec{y}}_{in} $ are input quadrature vectors over which the computation will be performed, and $\hat{\vec{X}}_{out}$, $ \hat{\vec{Y}}_{out} $  are output quadrature vectors obtained as a result of computation. In  Eq. (\ref{1}), the matrices $E$ and $ \tilde U$ depend on the cluster state configuration, because OWQC are implemented via local measurements of the entangled cluster state nodes.  It was originally believed that the richer the cluster state configuration (i.e., the more nodes and edges in the cluster state), the more transformations could be performed with it. However, this approach is only correct when using quantum oscillators with infinite squeezing, which, of course, is not realistic. If we talk about physical systems, then an increase in the number of nodes leads to an increase in errors associated with the finiteness of oscillators squeezing \cite{Korolev}. As a result, this error can grow to the point that it will be impossible to correct, and it will ruin all computation. Thus, possessing the oscillators with some finite squeezing, we should optimize the configuration of the cluster state so that it remains suitable for universal Gaussian transformations and would provide the smallest computation errors.

In this work, we will solve two problems.   First, we will select from all possible cluster configurations those that allow the implementation of universal Gaussian transformations. To that end, we will classify cluster states, and for each class, we will find general expressions of the type (\ref{1}). Next, we will use the obtained equations to determine what transformations can be performed on these states and select the required configurations. It will be the content of this article. In the next publication, we will compare all these configurations and find those that give the least error in the computation.

\section{Classification of cluster states}\label{class}
The ultimate goal of this work is to identify the configuration of the cluster state that allows performing a universal set of Gaussian operations with the smallest error. To this end, we want to find a  formula relating the input fields quadratures over which the transformation is performed to the output ones (the computation result) for an arbitrary cluster. It turns out that in general case, writing such an expression is not possible. However, we can divide all possible cluster configurations into several groups and find the required relations for each type of cluster. Next, we will explain the principle of division used. The main argument in support of this approach is the possibility of constructing an analytical solution for each class of cluster states.

Before explaining our classification principle, let's recall how the process of one-way quantum computation on continuous variables cluster states occurs.  The whole process can be divided into three steps. In the beginning, an $n$ nodes cluster state corresponding to a certain graph is prepared. We will assume that the cluster state is formed from the squeezed states of quantum oscillators via linear transformations (the Bogolyubov transform).  Here we will not dwell on specific methods of clusters creation. The second step consists in mixing $ m $ input states, over which the transformations will be carried out, to some nodes of the cluster state using symmetric beam splitters. The last step is the process of local homodyne measurements. All $ m $ input states and $ n-m $ cluster state nodes are measured. The quadratures of the $m$ remaining unmeasured cluster nodes will be determined by the Eq. (\ref{1}).

\begin{figure}[H]\centering
\includegraphics[scale=0.18]{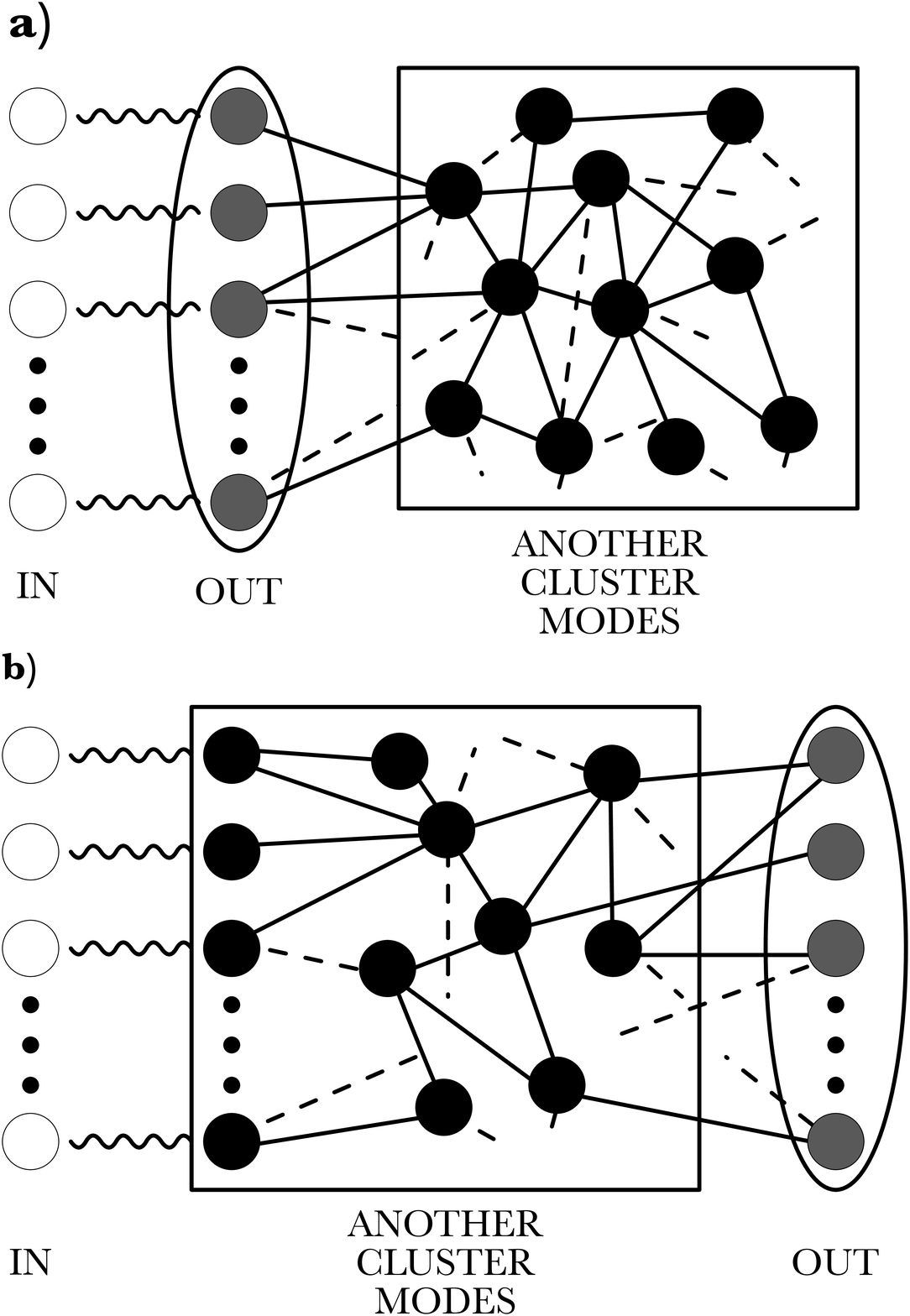}
\caption{Schematic of computation on cluster states:  IN is input modes, OUT is output modes, the wavy lines depict connections by symmetrical beam splitters. a) The input modes mix directly with the output ones. b) The input modes mix with measurable cluster modes. } \label{Fig_0}
\end{figure}

It can be seen from the described process that the nodes of the cluster to which the input modes are mixed are selected by the computation procedure itself. Therefore, it is essential whether local measurements are performed directly over these nodes or they are not measured and serve as output modes. It turns out that the two situations are mathematically different so that we will divide all computations by this principle.  This classification principle is shown in Fig. \ref{Fig_0}.   Besides, if the input modes mix with the measurable, the situation will significantly depend on the number of the cluster state nodes.  More precisely, on the ratio between the input states number and the total number of cluster nodes. Accordingly, we will consider three cases of computation: 1) input modes mix with output; 2) input modes mix with measurable, $n \leqslant 2m$; 3) input modes mix with measurable, $n>2m$.
\noindent Let us pass to the subject of finding relations between the output and input quadratures obtained by computation on cluster states of different configurations.

\section{Case of computation when the input states mix with output.} \label{section_n>m}

Consider the first case of quantum computing on cluster states. Suppose we have $m$ input states and an $n$ nodes cluster state. For computation, the number of cluster nodes should be no less than the number of input modes, therefore we will assume $n=m+l$, where $l \geqslant 0$. For further calculations, we need a relation between the cluster state quadratures and the quadratures of the squeezed quantum oscillators which used to generate this cluster. This relationship is defined as follows \cite{Ukai_2}:
\begin{align} \label{1.1}
\hat{\vec{X}}+i\hat{\vec{Y}}&=U \left(\hat{\vec{x}}_s+i\hat{\vec{y}}_s\right)=\left( {I}_n+i{A}\right)\Re {U}
\left(\hat{\vec{x}}_s+i\hat{\vec{y}}_s\right) \nonumber \\
&\equiv \left( {I_n}+i{A}\right)\;\left(\hat{\vec{x}}_r+i \hat{\vec{y}}_r\right),
\end{align}
where $\hat{\vec{x}}_r  \equiv \Re {U} \,\hat{\vec{x}}_s $, $
\hat{\vec{y}}_r \equiv \Re {U} \,\hat{\vec{y}}_s $, $\hat{\vec{X}}$ and $\hat{\vec{Y}}$ are vectors of cluster state quadratures, $\hat{\vec{x}}_s$ and $\hat{\vec{y}}_s$ are the vectors consisting of squeezed quantum oscillators quadratures, $U$ is a unitary transformation that turns oscillators in the squeezed state into oscillators in a cluster state (Bogoliubov transformation), $I_n$ is the identity matrix of dimension $n$, $A$ is the adjacency matrix of the cluster state graph.  Here, the transformation $U$ is organized in such a way that the cluster state nullifiers tend to zero. In Eq. (\ref{1.1}), we introduced new vectors of quadrature  $\hat{\vec{x}}_r  \equiv \Re {U} \;\hat{\vec{x}}_s $ and $\hat{\vec{y}}_r \equiv \Re {U} \;\hat{\vec{y}}_s $, which consist only of combinations of stretched and squeezed quadratures, respectively.

We mix the $m$ cluster state modes with $m$ input quantum states by symmetric beam splitters.
The quadratures after beam splitter can be written as follows:
\begin{align}
\hat{\vec{x}}_{in}'+i\hat{\vec{y}}_{in}'=&\frac{1}{\sqrt{2}}\left(\hat{\vec{x}}_{in}+I_1'\hat{\vec{x}}_r-A_1\hat{\vec{y}}_r \right)\nonumber \\
&+\frac{i}{\sqrt{2}}\left(\hat{\vec{y}}_{in}+I_1'\hat{\vec{y}}_r+A_1\hat{\vec{x}}_r \right),\\
\hat{\vec{X}}_{m}'+i\hat{\vec{Y}}_{m}'=&\frac{1}{\sqrt{2}}\left(\hat{\vec{x}}_{in}-I_1'\hat{\vec{x}}_r+A_1\hat{\vec{y}}_r \right)\nonumber \\
&+\frac{i}{\sqrt{2}}\left(\hat{\vec{y}}_{in}-I_1'\hat{\vec{y}}_r-A_1\hat{\vec{x}}_r \right). \label{n>m_0}
\end{align}
where $\hat{\vec{X}}_m'$ and $\hat{\vec{Y}}_{m}'$  are vectors of quadratures of the first $m$ cluster modes after the beam splitters, $\hat{\vec{x}}'_{in} $ and $\hat{\vec{y}}'_{in}$  is a vector of the input state quadratures after the beam splitters. To write these equations we used a block partition of the matrices $A$ and $I_n$ in the form
\begin{align} \label{block_1}
&A=\begin{pmatrix}
A_{1}\\
A_{2}
\end{pmatrix}=\begin{pmatrix}
A_{11} & A_{12}\\
A_{12}^T & A_{22}
\end{pmatrix}, \\
&I_n=\begin{pmatrix}
I_{1}'\\
I_{2}'
\end{pmatrix}=\begin{pmatrix}
I_{m} & \mathds{O}_{m \times l}\\
\mathds{O}_{l \times m} & I_{l}
\end{pmatrix}, \label{block_2}
\end{align}
where $\lbrace A_{1}, I_1' \rbrace \subset M^{m \times (m+l)}$,
$\lbrace A_{2}, I_{2}' \rbrace \subset M^{l \times (m+l)}$,
$\lbrace A_{11}, I_{m} \rbrace \subset M^{m \times m}$,
$\lbrace A_{22}, I_{l} \rbrace \subset M^{l \times l}$,
$\lbrace A_{12}, \mathds{O}_{m \times l} \rbrace \subset M^{m \times l}$,
$\mathds{O}_{k\times t}$ is the zero matrix of of dimension $k\times t$. Hereinafter, the set of all l $q
\times p$ matrices over the field of real numbers is denoted $M^{q \times p}$.  It is important to note that the cluster state does not depend on the numbering of modes, but only on the entanglements between them. Thus, we can always number the mods so that all mods that mix with inputs have numbers from $1$ to $m$.

At the next stage of computations, it is necessary to make measurements over all modes available in the system, except outputs. In the present case, the output modes (not measurable modes) are given by Eq. (\ref{n>m_0}). All measurements will be carried out using balanced homodyne detectors. The amplitudes of quantum photocurrents obtained as a homodyne measurement result have the form:
\begin{align}
\begin{cases}
\begin{aligned}
\cos \mathbf  \Theta _{in} &\left( \hat{\vec{x}}_{in}+I_1'\hat{\vec{x}}_r-A_1\hat{\vec{y}}_r\right)\\
&+\sin \mathbf \Theta _{in}\left( \hat{\vec{y}}_{in}+I_1'\hat{\vec{y}}_r+A_1\hat{\vec{x}}_r \right) =\frac{\sqrt{2}}{\beta _0}\hat{ \vec i}_{in}\\
\cos \mathbf \Theta_1 &\left( I_2'\hat{\vec{x}}_r-A_2\hat{\vec{y}}_r\right)\\
&+\sin \mathbf \Theta_1 \left(I_2'\hat{\vec{y}}_r+A_2\hat{\vec{x}}_r \right)=\frac{1}{\beta _0}\hat{ \vec i}
\end{aligned}
\end{cases},
\end{align}
where $\beta _0 \in \mathds{R}$ is an amplitude of the local oscillators of homodyne detectors, $\mathbf{ \Theta} _{in}=\mathrm{diag} \left(\theta _{in,1}, \theta _{in,2}, \dots ,\theta _{in,m}\right)$ and $\mathbf{\Theta} _1=\mathrm{diag} \left(\theta _{1,1}, \theta _{1,2}, \dots ,\theta
_{1,l}\right)$  are a diagonal matrices consisting of phases of local oscillators used in the homodyne detection of each mode. Accordingly, $ \cos \mathbf {\Theta} $ and $ \sin \mathbf {\Theta} $ denote functions from the diagonal matrix. Using partitioning (\ref{block_1}) and (\ref{block_2}), we can rewrite this system in block-matrix form:
\begin{align} \label{n>m_1}
&\begin{pmatrix}
\cos \mathbf\Theta _{in} +\sin \mathbf\Theta _{in} A_{11} & \sin \mathbf\Theta _{in} A_{12}\\
\sin\mathbf \Theta _{1} A_{12}^T & \cos \mathbf\Theta _{1} +\sin \mathbf\Theta _{1} A_{22}
\end{pmatrix}\hat{\vec{x}}_r \nonumber\\
&=\begin{pmatrix}
\cos  \mathbf \Theta _{in} A_{11}-\sin  \mathbf \Theta _{in} & \cos  \mathbf \Theta _{in}A_{12}\\
\cos  \mathbf \Theta _{1} A_{12}^T & \cos  \mathbf \Theta _{1} A_{22}-\sin  \mathbf \Theta _{1}
\end{pmatrix}\hat{\vec y}_r \nonumber\\
&-\begin{pmatrix}
\cos \mathbf \Theta _{in}\hat{\vec{x}}_{in}+\sin  \mathbf \Theta _{in}\hat{\vec{y}}_{in}\\
\mathds{O}_{m\times 1}
\end{pmatrix}+\frac{1}{\beta _0}\begin{pmatrix}
{\sqrt{2}}\hat{\vec i}_{in}\\
\hat{\vec i}
\end{pmatrix}.
\end{align}
The last term on the right-hand side is responsible for the measured photocurrent. After the measurement, it will be a classical quantity, the value of which we know in each specific experiment. Accordingly, we can offset it by shifting the quadrature at the required value \cite{Ukai_2}. Technically, this allows us to put all photocurrents equal to zero, implying that these values will be compensated.

We now proceed to solve the Eq. (\ref{n>m_1}) with respect to the vector of unknown variables $ \hat {\vec {x}} _r $.  To solve, we need to invert the matrix
\begin{align}
M=\begin{pmatrix}
\cos \mathbf\Theta _{in} +\sin \mathbf\Theta _{in} A_{11} & \sin \mathbf\Theta _{in} A_{12}\\
\sin\mathbf \Theta _{1} A_{12}^T & \cos \mathbf\Theta _{1} +\sin \mathbf\Theta _{1} A_{22}
\end{pmatrix}.
\end{align}
The matrix $M$ will be invertible if one of the block matrices on its diagonal is invertible. We will give two solutions. Note that the solutions coincide in the case of the invertibility of both block matrices.

Suppose that the matrix $ \cos \mathbf \Theta_ {in} + \sin \mathbf \Theta _ {in} A_ {11} $ is nonsingular, then we use the first blockwise inversion formula (\ref{frob_1}) and get $ M^{-1}$. Next, substituting the resulting $M^{-1}$ in the Eq.  (\ref{n>m_1}), we find the vector $\hat{\vec{x}}_r$. Finally, substituting this vector in (\ref{n>m_0}), we obtain the final expression for the output quadratures:
\begin{widetext}
\begin{multline} \label{n_more_m_general}
\begin{pmatrix}
\hat{\vec{X}}_{out}\\
\hat{\vec{Y}}_{out}
\end{pmatrix}=\frac{1}{\sqrt{2}}\begin{pmatrix}
I_m+LQ^{-1}\cos \mathbf{\Theta}_{in} & L Q^{-1}\sin \mathbf{\Theta}_{in}\\
VQ^{-1}\cos \mathbf{\Theta}_{in} & I_m+VQ^{-1}\sin \mathbf{\Theta}_{in}
\end{pmatrix}\begin{pmatrix}
\hat{\vec{x}}_{in}\\
\hat{\vec{y}}_{in}
\end{pmatrix}+\\
+\frac{1}{\sqrt{2}}\begin{pmatrix}
I_m & \mathds{O}_{m\times m}\\
A_{11} & -I_m
\end{pmatrix}\begin{pmatrix}
Q^{-1}\sin \mathbf{\Theta}\\
I_m
\end{pmatrix}
\begin{pmatrix}
I_m+A_{11}^2+A_{12}M_{e1} & A_{11}A_{12}+A_{12}M_{e2}
\end{pmatrix}\hat{\vec{y}}_r,
\end{multline}
\end{widetext}
where the notation was introduced for 
$Q=\cos \mathbf\Theta _{in} +\sin \mathbf\Theta _{in} A_{11}$,
$L=I_m+Q^{-1}\sin \mathbf{\Theta}_{in}A_{12}H^{-1}\sin \mathbf{\Theta}_1A_{12}^T$,
$V=A_{11}L-A_{12}H^{-1}\sin \mathbf{\Theta}_1A_{12}^T$,
$H=\sin \mathbf \Theta _1 \Big( A_{22}-A_{12}^TQ^{-1}\sin \mathbf \Theta_{in} A_{12}\Big)+\cos \mathbf \Theta _1$,
$M_{e1}=H^{-1}\Big(\cos \mathbf{\Theta}_{1}A_{12}^T -\sin \mathbf{\Theta}_1A_{12}^TQ^{-1}\left(\cos \mathbf{\Theta}_{in}A_{11}-\sin
\mathbf{\Theta}_{in}\right)\Big)$
and $M_{e2}=H^{-1}\Big(\cos \mathbf{\Theta}_{1}A_{22}-\sin \mathbf{\Theta}_{1}\left( A_{12}^TQ^{-1}\cos \mathbf{\Theta}_{in}A_{12}+I_l\right)\Big)$.

Now suppose that the matrix $\cos \mathbf \Theta _{1} + \sin \mathbf \Theta _{1} A_{l2}$ is nonsingular, then we use another blockwise inversion formula (\ref{frob_2}) to obtain $ M^{-1} $.  Acting in the same way as in the previous case, we get the relation between input and output quadratures:
\begin{widetext}
\begin{multline}
 \label{n_more_m_general_2}
\begin{pmatrix}
\hat{\vec{X}}_{out}\\
\hat{\vec{Y}}_{out}
\end{pmatrix}=\frac{1}{\sqrt{2}}\begin{pmatrix}
I_m+K^{-1}\cos \mathbf{\Theta}_{in} & K^{-1}\sin \mathbf{\Theta}_{in}\\
PK^{-1}\cos \mathbf{\Theta}_{in} & I_m+PK^{-1}\sin \mathbf{\Theta}_{in}
\end{pmatrix}\begin{pmatrix}
\hat{\vec{x}}_{in}\\
\hat{\vec{y}}_{in}
\end{pmatrix}+\\
+\frac{1}{\sqrt{2}}\begin{pmatrix}
I_m & \mathds{O}_{m\times m}\\
P & -I_m
\end{pmatrix}\begin{pmatrix}
K^{-1}\sin \mathbf{\Theta}_{in}\\
I_m
\end{pmatrix}
\begin{pmatrix}
I_m+A_{11}^2+A_{12}M_{e3} & A_{11}A_{12}+A_{12}M_{e4}
\end{pmatrix}\hat{\vec{y}}_r,
\end{multline}
\end{widetext}
where $K=\cos \mathbf{\Theta}_{in}+\sin \mathbf{\Theta}_{in}P$,
$P=A_{11}-A_{12}D^{-1}\sin \mathbf{\Theta}_{1}A_{12}^T$,
$D=\cos \mathbf{\Theta}_{1}+\sin\mathbf{\Theta}_{1} A_{22}$,
$M_{e3}=D^{-1}\left(\cos \mathbf{\Theta}_{1}A_{12}^T-\sin \mathbf{\Theta}_{1}A_{12}^TA_{11}\right) $
and $M_{e4}=D^{-1}\left(\cos \mathbf{\Theta}_{1}A_{22}-\sin \mathbf{\Theta}_{1}\left(A_{12}^TA_{12}+I_l \right) \right)$.

Although the Eqs. (\ref{n_more_m_general}) and (\ref{n_more_m_general_2}) look cumbersome, it is clear from them that the transformation matrices of input quadratures to output ones (the first terms on the right-hand sides of Eqs.) depend only on the cluster state configuration (adjacency matrix $A$) and on the phases of local oscillators of homodyne detectors (matrix  $\mathbf \Theta_{in}$). The second term in both cases also depends on the cluster state generation method ($ \hat {\vec {y}} _ r = \Re U\, \hat {\vec {y}} _s $), i.e. on the choice of the unitary matrix $ U$  that turns oscillators in the squeezed state into oscillators in a cluster state \cite{Korolev}. In other words, in OWQC with continuous-variables, the cluster state  configuration affects the computation result, and the generation method of this state affects only the resulting error. This result is associated with the exclusion of the vector $ \hat {\vec {x}} _r = \Re U\, \hat {\vec {x}} _s $ from the equations, therefore it is valid for any case of OWQC with continuous-variables.

Note that to obtain a complete solution in the Eqs. (\ref{n_more_m_general}) and (\ref{n_more_m_general_2}) it is necessary to calculate $ Q^{- 1} $, $ H^{- 1} $ and $ K ^ {- 1} $, $D^{-1} $. For some cluster configurations, it can be done analytically, but in general it is not possible. Despite this, the implicit solution is sufficient for our further purposes.
\subsection{Quantum transformations implemented in this computation scheme}
As mentioned earlier, what we want to do is implement universal Gaussian transformations.  For these purposes, we need to demonstrate the feasibility of two types of operations: universal single-mode and arbitrary two-mode (we choose \emph {CZ} as the last operation). Let us discuss the possibility of implementing these transformations in the described OWQC scheme.

\subsubsection{Single-mode transformations}
If we find a transformation that can not be implemented in proposed computation scheme, that would mean that such a scheme is not universal.

Let us consider the opportunity of implementing the Fourier transform in this scheme. We will discuss the implementation of the $m$ -dimensional Fourier transform, which is defined by the following equation:
\begin{align}
\begin{pmatrix}
\hat{\vec{X}}_{out} \\
\hat{\vec{Y}}_{out}
\end{pmatrix}=\begin{pmatrix}
\mathds{O}_{m\times m} & -I_m \\
I_m & \mathds{O}_{m\times m}
\end{pmatrix} \begin{pmatrix}
\hat{\vec{x}}_{in} \\
\hat{{\vec y}}_{in}
\end{pmatrix}.
\end{align}
In the case of computations given by the Eq. (\ref{n_more_m_general}) we get the Fourier transform when the following equalities take place
\begin{align}
&LQ^{-1}\cos \mathbf{\Theta}_{in}=-I_m,\\
&LQ^{-1}\sin \mathbf{\Theta}_{in}=-\sqrt{2}I_m,\\
&VQ^{-1}\cos \mathbf{\Theta}_{in}=\sqrt{2}I_m,\\
&VQ^{-1}\sin \mathbf{\Theta}_{in}=-I_m.
\end{align}
The condition $\sqrt{2}\cos \mathbf{\Theta}_{in}=\sin \mathbf{\Theta}_{in}$ is a prerequisite for the Fourier transforms implementation as it implies from the first and second equations.  With this in mind, the other two Eqs. are rewritten as $VQ^{-1}\cos \mathbf{\Theta}_{in}=\sqrt{2}I_m$ and $VQ^{-1}\cos \mathbf{\Theta}_{in}=-{1}/{\sqrt{2}}I_m$.  The contradiction in this regard implies that the Fourier transform can not be performed.  For the computation given by the Eq. (\ref{n_more_m_general_2}), the proof looks similar. 

A setback in forming described computation scheme proves it does not avail for the realization of universal Gaussian transformations. We could stop at this, but, to complete the picture, we will figure out the properties of two-mode operations in this scheme.

\subsubsection{Two-mode transformation}

Consider the implementation of the two-mode operation \emph{CZ}, which relates the input quadratures to output ones by the following rule:
\begin{align}  \label{CZ}
\begin{pmatrix}
\hat{\vec{X}}_{out} \\
\hat{\vec{Y}}_{out}
\end{pmatrix}=\begin{pmatrix}
I_m & \mathds{O} \\
W & I_m
\end{pmatrix}\begin{pmatrix}
\hat{\vec{x}}_{in} \\
\hat{\vec{y}}_{in}
\end{pmatrix} \equiv CZ\left[W\right]\begin{pmatrix}
\hat{\vec{x}}_{in} \\
\hat{\vec{y}}_{in}
\end{pmatrix},
\end{align}
where $W$ is an arbitrary matrix with a zero main diagonal. If we use the Eq. (\ref{n_more_m_general}) to implement this transformation, we get a set of conditions
\begin{align}
&LQ^{-1}\sin \mathbf{\Theta}_{in}=\mathds{O}_{m\times m},\label{16}\\
&VQ^{-1}\cos \mathbf{\Theta}_{in}=\sqrt{2}W,\label{17}\\
&LQ^{-1}\cos \mathbf{\Theta}_{in}=(\sqrt{2}-1)I_m,\label{18}\\
&VQ^{-1}\sin \mathbf{\Theta}_{in}=(\sqrt{2}-1)I_m.\label{19}
\end{align}
From Eq. (\ref{16}) we get that either $\sin \mathbf{\Theta}_{in}=\mathds{O}_{m\times m}$ or $ L=\mathds{O}_{m\times m}$.  If the first is true, then the Eq (\ref{19}) will not be satisfied. In the case when $L=\mathds{O}_{m\times m}$ Eq. (\ref{18}) ceases to hold. This contradiction suggests the impossibility of realizing a pure \emph{CZ} transformation. The proof for the case of using the Eq. (\ref{n_more_m_general_2}) is similar.

Although the \emph{CZ} operation is not feasible in this computation scheme, a similar transformation can be performed. Suppose $\sin \mathbf{\Theta}_{in} =\mathds{O}_{m\times m}$, then Eqs. (\ref{n_more_m_general})  and  (\ref{n_more_m_general_2}) will match and take the form:
\begin{multline} \label{end_n_more_m}
\begin{pmatrix}
\hat{\vec{X}}_{out} \\
\hat{\vec{Y}}_{out}
\end{pmatrix}=S\left(-\frac{\ln 2}{2}I_m\right) CZ\left[ {P} \right]\begin{pmatrix}
\hat{\vec{x}}_{in} \\
\hat{\vec{y}}_{in}
\end{pmatrix}\\
-\frac{1}{\sqrt{2}} \begin{pmatrix}
\mathds{O}_{m\times m} & \mathds{O}_{m\times l}\\
I_m+A_{11}^2+A_{12}M_{e3} & A_{11}A_{12}+A_{12}M_{e4}
\end{pmatrix}\hat{\vec{y}}_r,
\end{multline}
where a multimode squeezing matrix $S$ is defined as:
\begin{align} \label{S}
&S(\mathbf{r})=\begin{pmatrix}
\exp \left(-\mathbf{r}\right) & \mathds{O}_{m\times m}\\
\mathds{O}_{m\times m} & \exp \left(\mathbf{r}\right)
\end{pmatrix} ,
\end{align}
for $\mathbf{r}=\mbox{diag}\left(r_1,r_2,\dots, r_m\right)$. I.e. this computation scheme allows us to perform the \emph{CZ} transformation simultaneously with the squeezing transformation.  Moreover, if the first $m$ cluster modes are not entangled with the other ones (i.e., with $ A_ {12} = \mathds {O} _ {m \times m} $), we will get a same type transformation, but with less error:
\begin{multline} \label{n=m_CZ}
\begin{pmatrix}
\hat{\vec{X}}_{out} \\
\hat{\vec{Y}}_{out}
\end{pmatrix}=S\left(-\frac{\ln 2}{2}I_m\right) CZ\left[ A_{11} \right]\begin{pmatrix}
\hat{\vec{x}}_{in} \\
\hat{\vec{y}}_{in}
\end{pmatrix}\\
-\frac{1}{\sqrt{2}}\begin{pmatrix}
\mathds{O}_{m\times m} & \mathds{O}_{m\times l}\\
I_m+A_{11}^2 & \mathds{O}_{m\times l}
\end{pmatrix}\hat{\vec{y}}_r.
\end{multline}
Both the \emph {CZ} transformation and multimode squeezing belong to the Clifford group, therefore, if we add a universal single-mode operation to this computation scheme then, such a composite system can already be considered as a universal Gaussian transformer. Moreover, since the least possible amount of cluster nodes were used in this example,such a scheme is an eligible candidate for reducing computation errors. Moreover, since the least possible amount of cluster nodes were used in this example, such a scheme is an eligible candidate for reducing computation errors. So findings indicate that such an approach may be useful to the construction of computing schemes.

In this section, we have demonstrated that the computations in which the input states are mixed directly with the output ones are not universal. With such computations, it is impossible to implement either universal single-mode transformations or the pure \emph{CZ} transformation.

\section{Case of computation when the input states mix with measurable cluster nodes ($n \leqslant 2m$)} \label{sec_n=2m}
The previous section demonstrated the case of computation when the input modes mix with the output ones.  Hereafter, we will consider the opposite situation when the input modes mix with intermediate cluster ones, which will subsequently be measured. As we said in the section \ref{class}, we will divide all cases of such computations into two: $n\leqslant 2m$ and $n>2m$. In this section, we begin to consider the case when the number of cluster nodes exactly twice the amount of input modes ($ n = 2m $). Let us divide the vector of the cluster state quadratures into two components. The first consists of quadratures, which will be mixed with input states. The second consists of output quadratures.
\begin{multline}
\hat{\vec{X}}+i\hat{\vec{Y}}=\left(I_{n}+iA \right) \left(\hat{\vec{x}}_r+i\hat{\vec{y}}_r \right)\\
=\begin{pmatrix}
I_1'+iA_1 \\
I_2'+iA_2
\end{pmatrix}\left(\hat{\vec{x}}_r+i\hat{\vec{y}}_r \right)=\begin{pmatrix}
\hat{\vec{X}}_1+i\hat{\vec{Y}}_1\\
\hat{\vec{X}}_2+i\hat{\vec{Y}}_2
\end{pmatrix},
\end{multline}
where $\lbrace I_1'=\begin{pmatrix}
I_m & \mathds{O}_{m\times m}
\end{pmatrix}, I_2'=\begin{pmatrix}
\mathds{O}_{m\times m} & I_m
\end{pmatrix} \rbrace \subset M^{m \times 2m}$ are block matrices making up the identity matrix $I_{n} \in M^{2m \times 2m}$,
and $\lbrace A_1=\begin{pmatrix}
A_{11} & A_{12}
\end{pmatrix}, A_2=\begin{pmatrix}
A_{12}^T & A_{22}
\end{pmatrix} \rbrace \subset M^{m \times 2m}$  are adjacency matrix $A \in M^{2m \times 2m}$ blocks.

As in the previous case, the input modes defined by the vector $ \hat {\vec {x}} _ {in} + i \hat {\vec {y}}_{in} $ will be mixed with the first $m$ modes of the cluster state using symmetrical beam splitters. The modes obtained after the beam splitters are measured using balanced homodyne detectors. The results of such measurements can be written as a system of equations:
\begin{align} \label{sys_2m_1}
\begin{cases}
\begin{aligned}
\cos \mathbf {\Theta} _{in} &\left(\hat{\vec{x}}_{in}+I_1'\hat{\vec{x}}_r-A_1\hat{\vec{y}}_r \right)\\
&+\sin \mathbf {\Theta} _{in} \left(\hat{\vec{y}}_{in}+A_1\hat{\vec{x}}_r+I_1'\hat{\vec{y}}_r \right)=\frac{\sqrt{2}}{\beta_0}\hat{\vec{i}}_{in}\\
\cos \mathbf {\Theta} & \left(\hat{\vec{x}}_{in}-I_1'\hat{\vec{x}}_r+A_1\hat{\vec{y}}_r \right)\\
&+\sin \mathbf {\Theta}  \left( \hat{\vec{y}}_{in}-A_1\hat{\vec{x}}_r-I_1'\hat{\vec{y}}_r \right)=\frac{\sqrt{2}}{\beta_0}\hat{\vec{i}}
\end{aligned}
\end{cases}
\end{align}
where $ \mathbf {\Theta}_{in}=\mathrm{diag} \left( \theta _{in,1} ,\dots ,  \theta _{in,m}\right) $, $\mathbf {\Theta}=\mathrm{diag} \left( \theta _{1} ,\dots , \theta _{m}\right)$ are diagonal matrices consisting of local oscillators phases, $\beta _0 $ are  amplitudes of local oscillators. To obtain the relation between the vector of output quadratures and the input ones, we need to solve the system of equations (\ref{sys_2m_1}) and substitute the found vector $\hat{\vec{x}}_r$ in the unmeasured cluster modes $\hat{\vec{X}}_2+i\hat{\vec{Y}}_2$ (see Appendix \ref{sec_sol}). As a result we get the following equation:
\begin{multline} \label{end_n=2m}
\begin{pmatrix}
\hat{\vec{X}}_{out}\\
\hat{\vec{Y}}_{out}
\end{pmatrix}=CZ\left[A_{22}\right]\begin{pmatrix}
-A_{12} ^{-1} &\mathds{O}_{m\times m}\\
\mathds{O}_{m\times m} & -A_{12}^T
\end{pmatrix}R\left( -\frac{\pi}{2}I_m\right)\\
*CZ\left[A_{11}\right]R\left(\frac{\pi}{2} I_m-\frac{1}{2}\mathbf{\Theta}_+\right)S\left(\ln\left[ \tan \frac{1}{2} \mathbf{\Theta}_-\right]\right)\\
*R\left(-\frac{1}{2}\mathbf{\Theta}_+\right)\begin{pmatrix}
\hat{\vec{x}}_{in}\\
\hat{\vec{y}}_{in}
\end{pmatrix} -\begin{pmatrix}
A_{12}^{-1} & \mathds{O}_{m\times m}\\
A_{22}A_{12}^{-1} & -I_m
\end{pmatrix}\hat{\vec{N}},
\end{multline}
where $\mathbf {\Theta} _{\pm}=\mathbf {\Theta} \pm \mathbf {\Theta}_{in}$, $CZ\left[W\right]$ is the \emph{CZ} transformation matrix  defined by Eq. (\ref{CZ}), $S({r})$ is the multimode squeezing matrix defined by Eq. (\ref{S}), matrices $R\left(\mathbf{\Theta}\right)$ are multimode rotation matrices:
\begin{align} \label{R}
R\left(\mathbf{\Theta}\right)=\begin{pmatrix}
\cos \mathbf{\Theta} & -\sin \mathbf{\Theta}\\
\sin \mathbf{\Theta} & \cos \mathbf{\Theta}
\end{pmatrix}.
\end{align}
In Eq. (\ref{end_n=2m}), we neglected all the photocurrent amplitudes for the reasons described above. In addition, we used the relationship between the squeezed quadratures $ \hat {\vec{y}}_s$ and cluster state nullifiers \cite{Korolev} $\hat{\vec{N}}=\left(A^2+I_n \right) \Re U \hat{\vec{y}}_s$.

\subsection{Quantum transformations implemented in this computation scheme}
Let us now look at the transformations that can be performed in this computation scheme.

\subsubsection{CZ transformation}

From Eq. (\ref{end_n=2m}) we see, that if we put $A_{11}=\mathds{O}_{m\times m}$, $ \mathbf {\Theta}_+=\mathds{O}_{m\times m}$, $\mathbf{\Theta}_-=\frac{\pi}{2}I_m$ and $A_{12}=-I_m$, then we get the \emph{CZ} transformation. 
\begin{align} \label{n=2m_CZ}
&\begin{pmatrix}
\hat{\vec{X}}_{out}\\
\hat{\vec{Y}}_{out}
\end{pmatrix}=CZ\left[A_{22}\right]\begin{pmatrix}
\hat{\vec{x}}_{in}\\
\hat{\vec{y}}_{in}
\end{pmatrix} +\begin{pmatrix}
I_m & \mathds{O}_{m\times m}\\
A_{22} & I_m
\end{pmatrix}\hat{\vec{N}},
\end{align}
which is completely determined by the matrix $A_{22}$.
\subsubsection{Single-mode transformations}
Now consider the opportunity of implementing a universal single-mode transformation ($m=1$). The Bloch-Messiah reduction theorem \cite{Braunstein2005} states that a universal single-mode transformation matrix can be represented as a product $R\left(\varphi
_2\right)S\left(r\right)R\left(\varphi_1\right)$. In our case, provided $ A_ {22} = A_ {11} = 0 $ and $ A_ {12} = - 1 $  we get:
\begin{multline} \label{One_mode}
\begin{pmatrix}
\hat{{X}}_{out}\\
\hat{{Y}}_{out}
\end{pmatrix}=R\left(-\frac{\Theta_+}{2}\right)S\left(\ln\left[ \tan \frac{ \Theta _-}{2} \right]\right)\\
*R\left(-\frac{{\Theta}_+}{2}\right)\begin{pmatrix}
\hat{{x}}_{in}\\
\hat{{y}}_{in}
\end{pmatrix}+ \hat{\vec{N}}.
\end{multline}
We see that the transformation is not universal since the angles of the rotation matrices coincide.

Thus, in this scheme, the \emph{CZ} transform can be implemented, but it is impossible to obtain a universal single-mode transformation. However, we see that in the proposed calculation scheme, all Clifford group generators are realizable. This means that the universality of computations could be achieved through the repeated use of a computational procedure. To do this, we, of course, require more cluster modes.

Since the $ n = 2m $ cluster modes are not sufficient for universal computation,  it is clear that the case of $n <2m $ also will not lead to positive results and does not require additional consideration. Moreover, since the number of output modes should be equal to $m$, in this case, we will not have enough measurable cluster modes, and we will be forced to mix more than one input mode with some cluster modes. This will lead to the loss of information from some input modes, and the impossibility of OWQC. From a mathematical point of view, this result is obtained due to the presence of linearly dependent rows in the system of equations on $\hat {\vec{x}}_r $. Thus, there is a condition on mixing of input modes with the cluster: "each input mode should be associated with only one cluster mode" \;.

 \section{Case of computation when the input states mix with measurable cluster nodes ($n > 2m$)} \label{section_n>2m}
 Let us now consider the case when the number of cluster state nodes more than twice the amount of input modes, i.e., $ n = 2m + l $, for $ l \geqslant 0 $. In this case, the vector consisting of cluster state quadratures will be defined as follows:
\begin{align}
\hat{\vec{X}}+i\hat{\vec{Y}}=\begin{pmatrix}
\hat{\vec{X}}_{1}+i\hat{\vec{Y}}_{1}\\
\hat{\vec{X}}_{2}+i\hat{\vec{Y}}_{2}\\
\hat{\vec{X}}_{3}+i\hat{\vec{Y}}_{3}
\end{pmatrix}
=\begin{pmatrix}
I_{1}'+iA_{1} \\
I_{2}'+iA_{2} \\
I_3'+iA_3
\end{pmatrix} (\hat{\vec{x}}_r+i\hat{\vec{y}}_r),
\end{align}
where
 $I_{1}'=\begin{pmatrix}
I_{m} & \mathds{O}_{m\times m} & \mathds{O}_{m\times l}
\end{pmatrix}$,
$I_{2}'=\begin{pmatrix}
 \mathds{O}_{m\times m} & I_{m} & \mathds{O}_{m\times l}
\end{pmatrix}$,
$I'_3=\begin{pmatrix}
\mathds{O}_{l\times m} & \mathds{O}_{l\times m} & I_l
\end{pmatrix}$,
$A_{1}=\begin{pmatrix}
A_{11} & A_{12} & A_{13}
\end{pmatrix}$,
$A_{2}=\begin{pmatrix}
A_{12}^T & A_{22} & A_{23}
\end{pmatrix}$,
$A_{3}=\begin{pmatrix}
A_{13}^T & A_{23}^T & A_{33}
\end{pmatrix}$
 and
  $\lbrace A_{11},A_{12}, A_{22} \rbrace\subset M^{m\times m}$,
   $\lbrace A_{23}, A_{13}\rbrace   \subset M^{m\times l}$, 
   $A_{33} \in M^{l\times l}$.  Since the cluster state does not depend on the mode numbering, we have chosen here the order of the modes such that the quadratures with indices $1$ and $3$ are measurable, and the quadratures with index $2$ are output.

 We  mix the $\hat{\vec {X}} _ {1} + i \hat{\vec {Y}} _ {1}$ with the input modes $ \hat {\vec {x}}_{in} + i \hat {\vec{y}}_{in} $ and measure the resulting modes and modes $ \hat {\vec{X}}_{3} + i \hat {\vec{Y}}_{3}$. Let us write the photocurrent operators in the form:
\begin{align}
\begin{cases}
\begin{aligned}
\cos \mathbf{\Theta}_1&\left(\hat{\vec{x}}_{in}+I_{1}'\hat{\vec{x}}_r-A_{1}\hat{\vec{y}}_r \right)\\
&+\sin \mathbf{\Theta}_1\left(\hat{\vec{y}}_{in}+I_{1}'\hat{\vec{y}}_r+A_{1}\hat{\vec{x}}_r \right)=\frac{\sqrt{2}}{\beta _0}\hat{\vec{i}}_1\\
\cos \mathbf{\Theta}_2&\left(\hat{\vec{x}}_{in}-I_{1}'\hat{\vec{x}}_r+A_{1}\hat{\vec{y}}_r \right)\\
&+\sin \mathbf{\Theta}_2\left(\hat{\vec{y}}_{in}-I_{1}'\hat{\vec{y}}_r-A_{1}\hat{\vec{x}}_r \right)=\frac{\sqrt{2}}{\beta _0}\hat{\vec{i}}_2\\
\cos \mathbf{\Theta}_3&\left(I_{3}'\hat{\vec{x}}_r-A_{3}\hat{\vec{y}}_r \right)\\
&+\sin \mathbf{\Theta}_3\left(I_{3}'\hat{\vec{y}}_r+A_{3}\hat{\vec{x}}_r \right)=\frac{1}{\beta _0}\hat{\vec{i}}_3
\end{aligned}
\end{cases}
\end{align}
To solve this system, we need to invert the matrix
\begin{widetext}
\begin{align}
M=\begin{pmatrix}
\cos \mathbf{\Theta}_1+\sin \mathbf{\Theta}_1A_{11} & \sin \mathbf{\Theta}_1A_{12} & \sin \mathbf{\Theta}_1A_{13} \\
-\cos \mathbf{\Theta}_2-\sin \mathbf{\Theta}_2A_{11} & -\sin \mathbf{\Theta}_2A_{12} & -\sin \mathbf{\Theta}_2A_{13} \\
\sin \mathbf{\Theta}_3 A_{13}^T & \sin \mathbf{\Theta}_3 A_{23}^T & \cos \mathbf{\Theta}_3 +\sin \mathbf{\Theta}_3 A_{33}
\end{pmatrix} \equiv\begin{pmatrix}
\tilde Q & \tilde T \\
\tilde C & \tilde D
\end{pmatrix} ,
\end{align}
\end{widetext}
where we introduced the following notation
\begin{align*}
&\tilde Q=\begin{pmatrix}
\cos \mathbf{\Theta}_1+\sin \mathbf{\Theta}_1A_{11} & \sin \mathbf{\Theta}_1A_{12}  \\
-\cos \mathbf{\Theta}_2-\sin \mathbf{\Theta}_2A_{11} & -\sin \mathbf{\Theta}_2A_{12}
\end{pmatrix}\in M^{2m \times 2m},\\
&\tilde T=\begin{pmatrix}
\sin \mathbf{\Theta}_1A_{13} \\
-\sin \mathbf{\Theta}_2A_{13}
\end{pmatrix}\in M^{2m\times l},\\
&\tilde C=\begin{pmatrix}
\sin \mathbf{\Theta}_3 A_{13}^T & \sin \mathbf{\Theta}_3 A_{23}^T
\end{pmatrix} \in M^{l\times 2m},\\
&\tilde D =
\cos \mathbf{\Theta}_3 +\sin \mathbf{\Theta}_3 A_{33} \in M^{l \times l}.
\end{align*}
As before, depending on the invertibility of the blocks $\tilde Q$ and $\tilde D$, there are two blockwise inversion formulas.

Let the matrix $\tilde{Q}$ be invertible. In this case, we invert the matrix $M$ using the first blockwise inversion formula (\ref{frob_1}) and find the vector $\hat {\vec{x}}_r $. We substitute this vector into the output quadratures $\hat {\vec {X}}_{m_2} + i \hat{\vec{Y}}_{m_2}$. As a result, we can record the final relationship of the output quadrature with the input ones:
\begin{widetext}
\begin{multline} \label{n>2m_1}
\begin{pmatrix}
\hat{\vec{X}}_{out}\\
\hat{\vec{Y}}_{out}
\end{pmatrix}=CZ\left[A_{22}\right]\begin{pmatrix}
-(K_1A_{23}^T+I_m)A_{12}^{-1} & K_1K_2+A_{12}^{-1}A_{11}\\
A_{23} \tilde {H}^{-1}\sin \mathbf{\Theta}_3 A_{23}^TA_{12}^{-1} & -A_{12}^T-A_{23}\tilde H^{-1}\sin \mathbf{\Theta}_3K_2
\end{pmatrix} \\
*R\left(-\frac{1}{2}\mathbf{\Theta}_+\right)S\left(\ln\left[ \tan \frac{1}{2} \mathbf{\Theta}_-\right]\right)R\left(-\frac{1}{2}\mathbf{\Theta}_+\right)\begin{pmatrix}
\hat{\vec{x}}_{in}\\
\hat{\vec{y}}_{in}
\end{pmatrix}+E\hat{\vec{y}}_r,
\end{multline}
\end{widetext}
where $K_1=A_{12} ^{-1}A_{13}\tilde{H}^{-1}\sin \mathbf{\Theta}_3$, $K_2=A_{23}^TA_{12}^{-1}A_{11}-A_{13}^T$ and $\tilde H=\cos
\mathbf{\Theta}_3 +\sin \mathbf{\Theta}_3 \left( A_{33}- A_{23}^T  A_{12} ^{-1}A_{13}\right)$,
$\mathbf{\Theta}_{\pm}=\mathbf{\Theta}_1\pm\mathbf{\Theta}_2$,  matrices  ${CZ}\left[ W\right]$, $S(r)$ and $R (\mathbf{\Theta})$ are determined by Eqs. (\ref{CZ}), (\ref{S}) and (\ref{R}) respectively. An error matrix $E$ is written out in the Appendix \ref{Er_1}.

Assume the matrix $\tilde{D}$ is invertible. We find a solution for $\hat{\vec{x}}_r$  using  another blockwise inversion formula  (\ref{frob_2}) and again substitute it into $\hat{\vec{X}}_{m_2}+i\hat{\vec{Y}}_{m_2}$. As a result, the equation for the relationship between input and output quadrature is written as:
\begin{widetext}
\begin{multline} \label{n>2m_2}
\begin{pmatrix}
\hat{\vec{X}}_{out}\\
\hat{\vec{Y}}_{out}
\end{pmatrix}=CZ\left[A_{22}\right]\begin{pmatrix}
-\tilde  K_2^{-1} & \tilde K^{-1}_2 \tilde{K_1}\\
A_{23}\tilde{D}^{-1}\sin \mathbf{\Theta}_3 A_{23}^T\tilde K_{2}^{-1} & -A_{12}-A_{23}\tilde{D}^{-1}\sin \mathbf{\Theta}_3\tilde K_3
\end{pmatrix} \\
*R\left(-\frac{1}{2}\mathbf{\Theta}_+\right)S\left(\ln\left[ \tan \frac{1}{2} \mathbf{\Theta}_-\right]\right)R\left(-\frac{1}{2}\mathbf{\Theta}_+\right)\begin{pmatrix}
\hat{\vec{x}}_{in}\\
\hat{\vec{y}}_{in}
\end{pmatrix}+\tilde{E}\hat{\vec{y}}_r,
\end{multline}
\end{widetext}
where $\tilde{K}_1=A_{11}-A_{13}\tilde{D}^{-1}\sin \mathbf{\Theta}_3 A_{13}^T$, $\tilde{K}_2=A_{12}-A_{13}\tilde{D}^{-1}\sin \mathbf{\Theta}_3
A_{23}^T$, $\tilde{K}_3=A_{23}^T\tilde{K}_2^{-1} \tilde{K}_1-A_{13}^T$. Elements of the error matrix $\tilde E$ are written out in the Appendix \ref{Er_2}.

In this case of calculations (n> 2m), we obtained two relationships between the inputs and the outputs. The resulting equations are written in an implicit form, and for completeness, it is necessary to invert some matrices. This can be done numerically for clusters of a specific configurations.
\subsection{Quantum transformations implemented in this computation scheme}
Let us consider the transformations that can be performed in this computation scheme. First of all, we note that both Eqs. (\ref{n>2m_1}) and (\ref{n>2m_2}) are reduced to the Eq. (\ref{end_n=2m}) obtained in the case when the number of cluster modes exactly twice the number of input modes ($n=2m$). Formally, this is achieved by the absence of edges between the first $2m$ nodes of the cluster graph and the remaining $l$ nodes (i. e., when $A_{13} =A_{23} = \mathds{O}_{m \times l}$) .  Physically, this means that all transformations that we can perform in a scheme with fewer nodes, we can implement here. For example, the CZ transformation. Moreover, to implement this transformation, it is better to use a scheme with $n=2m$, since a smaller number of nodes provides fewer sources of error associated with the finite squeezing of the used oscillators. Now let us consider single-mode transformations that cannot be implemented in the case of $n=2m$.
\subsubsection{Single-mode transformations}

It would be desirable to perform single-mode transformations using the minimum number of cluster modes since each additional mode adds an error to the computation result. We know that it is impossible to implement a universal single-mode transfotmations on the two-node cluster state (see (\ref{One_mode})), so let's look at a three-node cluster state, $n=3$. In this case Eqs. (\ref{n>2m_1}) and (\ref{n>2m_2}) coincide and are equal to
\begin{multline} \label{n=3}
\begin{pmatrix}
\hat{{X}}_{out}\\
\hat{{Y}}_{out}
\end{pmatrix}=\frac{1}{d}\begin{pmatrix}
-\cos \Theta_3 &  -a_{13}^2\sin \Theta_3\\
a_{23}^2 \sin \Theta_3 & -a_{12}d+a_{13}a_{12}a_{23}\sin \Theta_3
\end{pmatrix}\\
*R\left(-\frac{1}{2}{\Theta}_+\right)S\left(\ln\left[ \tan \frac{1}{2} {\Theta}_-\right]\right)R\left(-\frac{1}{2}{\Theta}_+\right)
\begin{pmatrix}
\hat{{x}}_{in}\\
\hat{{y}}_{in}
\end{pmatrix}+{E}\hat{\vec{y}}_r,
\end{multline}
where $a_{ij}$  are the weights of the three-node graph, and $d=a_{12}\cos \Theta_3 -a_{23}a_{13}\sin \Theta_3$. The matrix of the resulting transformation is not a universal symplectic matrix. This can be proved as follows. Suppose we want to be able to implement a class of symplectic transformations defined by matrices
\begin{align} \label{n=3m_2}
\begin{pmatrix}
z \cos \Theta +(z+1)  \sin \Theta & (z+1) \cos \Theta - z \sin\Theta \\
- \cos \Theta - \sin \Theta & -\cos \Theta + \sin \Theta
\end{pmatrix},
\end{align}
for various $z \in \mathds{N}$. The transformation matrix in the Eq. (\ref{n=3}) coincides with the matrix (\ref{n=3m_2}) when
\begin{align}
&a_{13} = a_{23} \sqrt{1 + z}, \quad a_{12} = \frac{1}{1 + \sqrt{1 + z}},\\
&\Theta_3 = \rm arccot \left( a_{23}^2 z\right),\quad \Theta_-={\pi}/{2}, \quad  \Theta_+=-\Theta.
\end{align}
From these conditions, it is clear that for the implementation of various transformations  (\ref{n=3m_2}) (for different values of $z$), we should each time change the weight coefficients of the cluster state. Since the weight coefficients are determined by a cluster state generation method,  it will be necessary to generate new cluster states. This is difficult from a practical point of view. As a result, we can conclude that three-node cluster states are not suitable for implementing universal single-mode transformations.

\begin{figure}[H]
\centering
\includegraphics[scale=1.18]{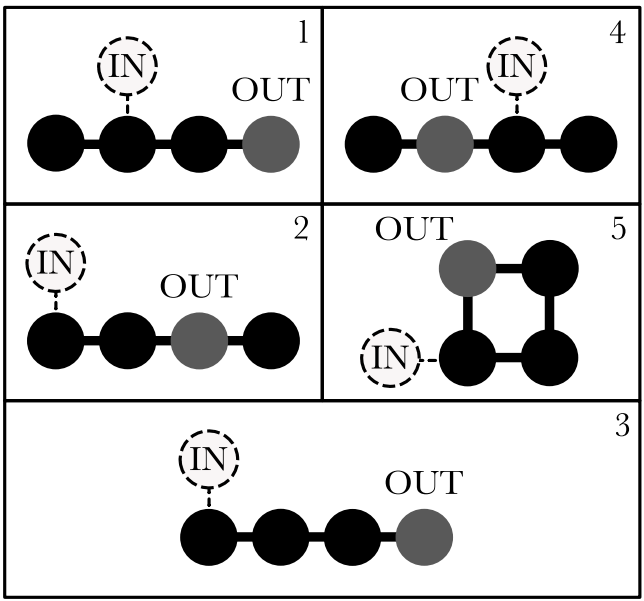}
\caption{Five types of cluster state configurations suitable for implementing universal single-mode transformations. In the figure, the dashed lines indicate the input states IN, which will be mixed with the cluster state nodes; OUT is the output mode.} \label{Fig_1}
\end{figure}

Let us now consider the case when $n=4$. To find cluster state configurations on which universal single-mode Gaussian transformations can be implemented, we iterate over the adjacency matrix elements of the four-node graph.  As the coefficients $a_{ij}$ in Eqs. (\ref{n>2m_1}) and (\ref{n>2m_2}), the numbers 0 or 1 are used (unweighted graph). After studying all possible cases, we get five types of different cluster state configurations of suitable for universal single-mode transformations. All these configurations are shown in Fig. \ref{Fig_1}.
\noindent   The output quadratures, obtained in the computation on the clusters of the presented configurations, are related to the input quadratures as follows
\begin{align}
\begin{pmatrix}
\hat{X}_{out}\\
\hat{Y}_{out}
\end{pmatrix}=\tilde{U}_j\begin{pmatrix}
\hat{x}_{in}\\
\hat{y}_{in}
\end{pmatrix}+E_j\hat{\vec{y}}_r, \qquad j=1,2,\dots,5,
\end{align}
where transformation matrices have the form:
\begin{align}
\tilde{U}_1&=\begin{pmatrix} \label{41}
 \cot \Theta _{4} \tan \Theta _{3}-1 & \cot \Theta _{4} \\
 -\tan \Theta _{3} & -1 \\
\end{pmatrix} R\left(-\frac{\pi}{2}\right) \nonumber \\
*&R\left(-\frac{1}{2}\Theta_+\right)S\left(\ln \left[\tan \frac{1}{2}\Theta_-\right]\right)R\left(-\frac{1}{2}\Theta_+\right),\\
\tilde{U}_2&=R\left(-\frac{\pi}{2} \right)\begin{pmatrix}
 \cot \Theta _{3} \tan\Theta _{4}-1 & \tan \Theta
   _{4} \\
 -\cot \Theta _{3} & -1 \\
\end{pmatrix} \nonumber\\
&*R\left(-\frac{1}{2}\Theta_+\right)S\left(\ln \left[\tan \frac{1}{2}\Theta_-\right]\right)R\left(-\frac{1}{2}\Theta_+\right),\\
\tilde{U}_3&=\begin{pmatrix}
\cot\Theta_{4} \cot \Theta_{3}-1  &  \cot \Theta_{4}\\
-\cot \Theta_{3} & -1
\end{pmatrix} \nonumber\\
&*R\left(-\frac{1}{2}\Theta_+\right)S\left(\ln \left[\tan \frac{1}{2}\Theta_-\right]\right)R\left(-\frac{1}{2}\Theta_+\right),\\
\tilde{U}_4&=R\left(-\frac{\pi}{2} \right)\begin{pmatrix}
\tan \Theta _{3}\tan \Theta _{4}-1 & \tan \Theta _{4} \\
 -\tan \Theta _{3} & -1
\end{pmatrix} R\left(-\frac{\pi}{2}\right)  \nonumber \\
&*R\left(-\frac{1}{2}\Theta_+\right)S\left(\ln \left[\tan \frac{1}{2}\Theta_-\right]\right)R\left(-\frac{1}{2}\Theta_+\right) ,\\
\tilde{U}_5&=\begin{pmatrix} \label{46}
 \tan \Theta_{ 3} \tan \Theta_{4}-1 & \tan \Theta_{4} \\
 -\tan \Theta_{ 3} & -1
\end{pmatrix} \nonumber \\
&*R\left(-\frac{1}{2}\Theta_+\right)S\left(\ln \left[\tan \frac{1}{2}\Theta_-\right]\right)R\left(-\frac{1}{2}\Theta_+\right),
\end{align}
and error matrices $E_j$ are written out in the Appendix \ref{sec_er_4}. In \cite{Ukai}, using additional lemmas and theorems, it was shown that these matrices are universal single-mode Gaussian transformation matrices. We have found explicit decomposition of these matrices as a product of $R\left (\varphi_1\right) S\left(r\right)R\left(\varphi_2-\Theta_+\right)$ (see Appendix \ref{function}), which is a simpler and shorter proof of the universality of matrices (\ref{41})-(\ref{46}).

Thus, we obtained that universal single-mode transformations are implemented using the five types of four-mode cluster states. There are no other suitable four-node clusters with an unweighted graph. There is no point in considering configurations with a more number of nodes since the non-ideal squeezing of each node is an additional source of errors. So we found the optimal configurations since they have a minimum number of nodes and can be used for universal computing.

\section{Conclusion}
In this paper, we analyzed all possible configurations of cluster states to identify those that allow the implementation of universal Gaussian transformations. In doing so, we have in mind the finding optimal cluster configurations that also provide a minimal computation error. In the next publication, we will focus on the analysis of errors in the found configurations.

The solution to this problem required us to introduce a classification: we divided all possible cluster states by the method of mixing inputs modes and by the number of nodes. In the first type of classification, we included all cases of computations in which the input states mix directly with the output ones.  One must admit that in these cases, it is possible to implement two-mode transformations (for example, \emph{CZ}), but it is impossible to perform single-mode operations. Therefore, such schemes are not universal. To another type of classification, we referred all the computations in which the input states mix with the measurable cluster modes. We have shown that if the number of cluster state nodes more than twice the amount of input modes, then computation on such clusters can be universal. If the number of cluster nodes exactly twice the amount of input modes, then the transformations implemented on these clusters will be generators of the Clifford group. This means that any operations can be implemented in these schemes, but for this, it is necessary to repeatedly send the result of some transformations to the input of others.

To complete the picture,  it would be necessary to consider another case of computation, when some input modes mix directly to the output cluster nodes, while others do through the measurable ones. However, after analyzing this case, we obtained a cumbersome solution, which is essentially a linear combination of the solutions obtained in the sections \ref{section_n>m} and \ref{section_n>2m}. This means that such type of computation is not universal, because the transformation of input modes mixed directly to the output ones is not universal.
 
 In addition, the analysis of all obtained relationships, between the input and output quadratures, showed that in all cases the transformation itself depends only on the cluster state configuration, but not on the method of its generation. At the same time, the measurement error depends on both factors.
 
Thus, it was shown in the paper what type of cluster configurations can be suitable for universal Gaussian transformations. The next step in the development of this topic will be a comparison of these configurations with each other and to identify optimal for computations among them. The solution to this problem will help to take us one step closer to the practical realization of the universal quantum computer.

\section{Acknowledgment}
This work was supported by the Russian Foundation for Basic Research (Grant Nos 18-32-00255mol\_a, 19-02-00204a and 18-02-00648a) and as part of the research activities of the Center of Quantum Technologies, Moscow State University M.V. Lomonosov.

\appendix
\section{Blockwise inversion}
The first blockwise inversion formula is used when the matrix $Q$ is nonsingular
\begin{align} \label{frob_1}
\begin{pmatrix}
Q & T \\
C & D
\end{pmatrix} ^{-1}
=
\begin{pmatrix}
Q^{-1} +Q^{-1} TH^{-1} CQ^{-1}  & -Q^{-1} TH^{-1}  \\
         -H^{-1} CQ^{-1}  & H^{-1}
\end{pmatrix},
\end{align}
where $H=D-CQ^{-1} T $. The second blockwise inversion formula is used when the matrix $D$ is nonsingular
\begin{align}
\begin{pmatrix} \label{frob_2}
Q & T \\
C & D
\end{pmatrix}^{-1}= \begin{pmatrix} \Pi^{-1} & -\Pi^{-1}TD^{-1} \\ -{D}^{-1}{C}\Pi^{-1} & \quad  {D}^{-1}+{D}^{-1}{C}\Pi^{-1}T{D}^{-1}\end{pmatrix},
\end{align}
where $\Pi=Q-T{D}^{-1}{C}$.

\section{The solution of the system of equations for the computation case when the input states mix with the measurable cluster nodes  ($n=2m$)} \label{sec_sol}
The steps, described in \ref{sec_n=2m}, lead us to the relationships between input and output quadratures in the form:
\begin{multline} \label{mat_sol}
\begin{pmatrix}
\hat{\vec{X}}_{out}\\
\hat{\vec{Y}}_{out}
\end{pmatrix}=\begin{pmatrix}
-A_{12} ^{-1} &A_{12}^{-1}A_{11}\\
-A_{22}A^{-1}_{12} & A_{22}A^{-1}_{12}A_{11}-A_{12}^T 
\end{pmatrix}\\
*\Phi\left(\mathbf{\Theta}_+,\mathbf{\Theta}_-\right)\begin{pmatrix}
\hat{\vec{x}}_{in}\\
\hat{\vec{y}}_{in}
\end{pmatrix} -\begin{pmatrix}
A_{12}^{-1} & \mathds{O}_{m\times m}\\
A_{22}A_{12}^{-1} & -I_m
\end{pmatrix}\begin{pmatrix}
\hat{\vec{N}}_1\\
\hat{\vec{N}}_2
\end{pmatrix},
\end{multline}
where  \begin{multline}
\Phi\left(\mathbf{\Theta}_+,\mathbf{\Theta}_-\right)=\begin{pmatrix}
\cos \mathbf{\Theta}_{-}+\cos \mathbf{\Theta}_{+} & \sin \mathbf{\Theta}_{+} \\
-\sin \mathbf{\Theta}_{+} & \cos \mathbf{\Theta}_{+}-\cos\mathbf{\Theta}_{-}
\end{pmatrix}\\
*\begin{pmatrix}
\csc \mathbf{\Theta}_{-} & \mathds{O}_{m\times m}\\
\mathds{O}_{m\times m} &\csc \mathbf{\Theta}_{-}
\end{pmatrix}.
\end{multline}
In the Eq. (\ref{mat_sol}), we replaced the vector $\hat{\vec{y}}_r$ with the vector of nullifiers using the following formula
\begin{multline}
\hat{\vec{N}}=\left(A^2+I_n \right) \hat{\vec{y}}_r\equiv \begin{pmatrix}
\hat{\vec{N}}_1\\
\hat{\vec{N}}_2
\end{pmatrix}\\
=\begin{pmatrix}
A_{11}^2+A_{12}A_{12}^T+I_m & A_{11}A_{12}+A_{12}A_{22} \\
A_{12}^TA_{11}+A_{22}A_{12}^T & A_{12}^TA_{12}+A_{22}^2+I_m
\end{pmatrix}\hat{\vec{y}}_r.
\end{multline}
Let us decompose the transformation matrix in (\ref{mat_sol}) into the product of the elementary transformation matrices. We start with the matrix that depends only on the weight coefficients of the cluster state
\begin{multline}
\begin{pmatrix}
-A_{12} ^{-1} &A_{12}^{-1}A_{11}\\
-A_{22}A^{-1}_{12} & A_{22}A^{-1}_{12}A_{11}-A_{12}^T 
\end{pmatrix}\\
=\begin{pmatrix}
I_m & \mathds{O}_{m\times m}\\
A_{22} & I_m
\end{pmatrix}\begin{pmatrix}
-A_{12} ^{-1} &\mathds{O}_{m\times m}\\
\mathds{O}_{m\times m} & -A_{12}^T 
\end{pmatrix}\begin{pmatrix}
I_m & -A_{11}\\
\mathds{O}_{m\times m} & I_m 
\end{pmatrix}. 
\end{multline}
The Bloch-Messiah decomposition for matrix $\Phi \left(\mathbf{\Theta}_+, \mathbf{\Theta}_-\right)$ is written as follows
\begin{align}
\Phi\left(\mathbf{\Theta}_+,\mathbf{\Theta}_-\right)=R\left(-\frac{1}{2}\mathbf{\Theta}_+\right)S\left(\ln\left[ \tan \frac{1}{2} \mathbf{\Theta}_-\right]\right)R\left(-\frac{1}{2}\mathbf{\Theta}_+\right).
\end{align}
Using these decompositions, we get the Eq. (\ref{end_n=2m}).
\begin{widetext}
\section{Error matrix of the OWQC when $n> 2m$. The first case of inversion of the matrix M} \label{Er_1}
The error matrix $E$ has the following elements
\begin{align}
\left[E\right]_{11}=&-(K_1A_{23}^T+I_m)A_{12}^{-1}-\left(K_1K_2+A_{12}^{-1}A_{11}\right)A_{11}-A_{12}^{-1}A_{13}\tilde{H}^{-1}\cos \mathbf{\Theta}_3 A_{13}^T -A_{12}^T,\\
\left[E\right]_{12}=& -\left(K_1K_2+A_{12}^{-1}A_{11}\right)A_{12}-A_{12}^{-1}A_{13}\tilde{H}^{-1}\cos \mathbf{\Theta}_3 A_{23}^T -A_{22},\\
\left[E\right]_{13}=&-\left(K_1K_2+A_{12}^{-1}A_{11}\right)A_{13}+A_{12}^{-1}A_{13}\tilde{H}^{-1}\left( \sin \mathbf{\Theta}_3-\cos \mathbf{\Theta}_3 A_{33}\right) -A_{23},\\
\left[E\right]_{21}=&A_{12}^TA_{11}+ A_{23} \left(\tilde{H}^{-1}\sin \mathbf{\Theta}_3\left[A_{23}A_{12}^{-1}+K_2A_{11}\right]+\tilde{H}^{-1} \cos \mathbf{\Theta}_3A_{13}^T \right)\nonumber\\
&+A_{22}\left( -(K_1A_{l2}+I_m)A_{12}^{-1}-\left(K_1K_2+A_{12}^{-1}A_{11}\right)A_{11} -A_{12}^{-1}A_{13}\tilde{H}^{-1}\cos \mathbf{\Theta}_3 A_{13}^T\right),\\
\left[E\right]_{22}=&I_m+A_{12}^TA_{12}+A_{22}\left( -\left(K_1K_2+A_{12}^{-1}A_{11}\right)A_{12} -A_{12}^{-1}A_{13}\tilde{H}^{-1}\cos \mathbf{\Theta}_3 A_{23}^T\right) \nonumber\\
&+A_{23}\left(\tilde{H}^{-1}\sin \mathbf{\Theta}_3K_2A_{12}+\tilde{H}^{-1}\cos \mathbf{\Theta}_3 A_{23}^T\right),\\
\left[E\right]_{23}=&A_{12}^TA_{13}+A_{22}\left(-\left(K_1K_2+A_{12}^{-1}A_{11}\right)A_{13}+A_{12}^{-1}A_{13}\tilde{H}^{-1}\left( \sin \mathbf{\Theta}_3-\cos \mathbf{\Theta}_3 A_{33}\right)\right)\nonumber\\
&+A_{23}\left(\tilde{H}^{-1}\sin \mathbf{\Theta}_3K_2A_{13}+\tilde{H}^{-1}\left( \cos \mathbf{\Theta}_3 A_{33}-\sin \mathbf{\Theta}_3\right)\right).
\end{align}

 \section{Error matrix of the OWQC when $n> 2m$. The second case of inversion of the matrix M} \label{Er_2}
The error matrix $ E $ has the following elements
\begin{align}
\left[\tilde E\right]_{11}=&-\tilde K_2^{-1}\tilde{K}_1A_{11}-\tilde K_2^{-1}A_{13}\tilde{D}^{-1}\cos \mathbf{\Theta}_3A_{13}^T-\tilde K_2^{-1} -A_{12}^T,\\
\left[\tilde E\right]_{12}=&-\tilde K_2^{-1}\tilde{K}_1A_{12}-\tilde K_2^{-1}A_{13}\tilde{D}^{-1}\cos \mathbf{\Theta}_3A_{23}^T -A_{22},\\
\left[\tilde E\right]_{13}=&-\tilde K_2^{-1}\tilde{K}_1A_{13}+\tilde K_2^{-1}A_{13}\tilde{D}^{-1}\left(\sin \mathbf{\Theta}_3-\cos \mathbf{\Theta}_3A_{33}\right) -A_{23},\\
\left[\tilde E\right]_{21}=&A_{12}^TA_{11} -A_{22}\tilde K_2^{-1}\left(\tilde{K}_1A_{11}+A_{13}\tilde{D}^{-1}\cos \mathbf{\Theta}_3A_{13}^T+I_m\right)\nonumber\\
&+A_{23} \tilde{D}^{-1}\left(\sin \mathbf{\Theta}_3 A_{23}^T\tilde{K}_2-\sin \mathbf{\Theta}_3 \tilde{K}_3 A_{11}+\left(I_l+\sin \mathbf{\Theta}_3A_{23}^T\tilde K_2^{-1}A_{13}\tilde{D}^{-1}\right)\cos \mathbf{\Theta}_3A_{13}^T \right),\\
\left[\tilde E\right]_{22}=&I_m+A_{12}^TA_{12} -A_{22}\tilde K_2^{-1}\left(\tilde{K}_1A_{12}+A_{13}\tilde{D}^{-1}\cos \mathbf{\Theta}_3A_{23}^T\right)\nonumber\\
&-A_{23}\tilde{D}^{-1}\left(\sin \mathbf{\Theta}_3 \tilde{K}_3 A_{12}-\left(I_l+\sin \mathbf{\Theta}_3A_{23}^T\tilde K_2^{-1}A_{13}\tilde{D}^{-1}\right)\cos \mathbf{\Theta}_3A_{23}^T\right), \\
\left[\tilde E\right]_{23}=&A_{12}^TA_{13} +A_{22}\tilde K_2^{-1}\left(\tilde{K}_1A_{13}-A_{13}\tilde{D}^{-1}\left(\sin \mathbf{\Theta}_3-\cos \mathbf{\Theta}_3A_{33}\right),\right)\nonumber\\
&-A_{23} \tilde{D}^{-1}\left(\sin \mathbf{\Theta}_3 \tilde{K}_3 A_{13}+\left(I_l+\sin \mathbf{\Theta}_3A_{23}^T\tilde K_2^{-1}A_{13}\tilde{D}^{-1}\right)\left(\sin \mathbf{\Theta}_3-\cos \mathbf{\Theta}_3A_{33}\right)\right).
\end{align}

\section{Error matrices resulting from the implementation of single-mode transformations on four-node cluster states} \label{sec_er_4}
The error matrices obtained during the implementation of single-mode transformations on four-node cluster states have the form:
\begin{align}
&E_1=\begin{pmatrix}
 -3 \cot \Theta _{4} & -\cot \Theta _{4} & 1-2 \cot
  \Theta _{4} \tan \Theta _{3} & 3-\cot \Theta
   _{4} \tan \Theta _{3} \\
 -2 & 1 & -2 \tan \Theta _{3} & -\tan \Theta _{3}
\end{pmatrix}, \\
&E_2=\begin{pmatrix}
 -2 \cot  \Theta _{3}  & -\cot  \Theta _{3}  & 3 & 1 \\
 2 \cot  \Theta _{3}  \tan  \Theta _{4} -1 & \cot  \Theta
   _{3}  \tan  \Theta _{4} +2 & -2 \tan  \Theta _{4}  &
   \tan \Theta _{4}
\end{pmatrix},\\
&E_3=\begin{pmatrix}
2\cot \Theta_{3} \cot \Theta_{4}-1 & \cot \Theta_{3} & 2+\cot \Theta_{3} \cot \Theta_{4} & 3\cot \Theta_{3}\\
-2\cot \Theta_{4} & 1 & -\cot \Theta_{4} & -2
\end{pmatrix}, \\
&E_4=\begin{pmatrix}
 3 & \tan  \Theta _{3}  & 2 \tan  \Theta _{3}  & 1 \\
 -2 \tan  \Theta _{4}  & 3-\tan  \Theta _{3}  \tan  \Theta
   _{4}  & 1-2 \tan  \Theta _{3}  \tan  \Theta _{4}  &
   \tan \Theta _{4}
\end{pmatrix}, \\
&E_5=\begin{pmatrix}
 \tan \Theta_{ 3} \tan \Theta_{ 4}-3 & -2 \tan \Theta_{ 4} & -\tan \Theta_{ 3} \tan
   \Theta_{ 4}-2 & -3 \tan \Theta_{ 4} \\
 -\tan \Theta_{ 3} & 3 & \tan \Theta_{3} & 2
\end{pmatrix},
\end{align}
\end{widetext}
\section{Bloch-Messiah decomposition for the matrices (\ref{41})-(\ref{46}) } \label{function}
The Bloch-Messiah decomposition obtained by us for the matrices (\ref{41}) - (\ref{46}) is as follows:
\begin{align}
\tilde U_j=R\left(-\frac{\pi}{2}\cdot l+\varphi_1\right)S\left(r\right)R\left(\varphi_2-\frac{\pi}{2}\cdot p-\Theta_+\right),
\end{align} 
for $j=1,\dots ,5$. Here $l$ is equal to one for $j \in \lbrace 2,4 \rbrace$ and equal to zero otherwise;  $p$ is equal to one for $ j \in \lbrace 1,4 \rbrace $ and equal to zero for the remaining $j$. Such decompositions are obtained for $\Theta_- = \pi/2$ and
\begin{widetext}
\begin{align}
&\varphi_2=\arctan\left[\frac{r \left(\csc ^3\varphi _1 \sqrt{\sin ^2\varphi _1 \left(r^4 \cos ^2\varphi _1-r^2+\sin ^2\varphi _1\right)}-r \cot \varphi _1\right)}{r^2 \csc ^2\varphi _1-1}\right]\\
&\lbrace \tan \Theta_3, \cot\Theta_3\rbrace=\frac{\csc \varphi _1 \sqrt{\sin ^2\varphi _1 \left(r^4 \cos ^2\varphi _1-r^2+\sin ^2\varphi _1\right)}}{r}, \\
&\lbrace \tan \Theta_4, \cot\Theta_4\rbrace=\frac{2 \left(r^4-1\right) \cot \varphi _1+\sqrt{2} r \csc ^3\varphi _1 \sqrt{\left(r^2-1\right) \sin ^2 \varphi   _1 \left(\left(r^2+1\right) \cos \left(2 \varphi _1\right)+r^2-1\right)}}{2 r^4 \cot ^2\varphi _1+2},
\end{align}
\end{widetext}
where the choice of the function $\tan$ or $\cot$ depends on the transformation number $j$.
\end{document}